\documentclass{elsart}
\usepackage{natbib,epsfig}
\begin{document}
\runauthor{Lessard, Cay\'on, Sembroski and Gaidos}
\begin{frontmatter}
\title{Wavelet Imaging Cleaning Method for 
Atmospheric Cherenkov Telescopes}
\author[Purdue]{R.W. Lessard\thanksref{Canada}}
\author[IFCA]{L. Cay\'on}
\author[Purdue]{G.H. Sembroski}
\author[Purdue]{J.A. Gaidos}

\address[Purdue]{Department of Physics, Purdue University, 
		West Lafayette, IN, 47907, USA}
\thanks[Canada]{Current address - Merak Projects, Ltd., 600, 322 11th Ave. S.W., Calgary, Alberta, Canada, T2R 0C5}
\address[IFCA]{Instituto de F\'\i sica de Cantabria, Fac. Ciencias, 
avda. Los Castros s/n, 39005 Santander, Spain}

\begin{abstract}
We present a new method of image cleaning for imaging atmospheric
Cherenkov telescopes. The method is based on the utilization of
wavelets to identify noise pixels in images of gamma-ray and hadronic
induced air showers. This method selects more signal pixels with
Cherenkov photons than traditional image processing techniques. In
addition, the method is equally efficient at rejecting pixels with
noise alone. The inclusion of more signal pixels in an image of an air
shower allows for a more accurate reconstruction, especially at lower
gamma-ray energies that produce low levels of light. We present the
results of Monte Carlo simulations of gamma-ray and hadronic air
showers which show improved angular resolution using this cleaning
procedure. Data from the Whipple Observatory's 10-m telescope are
utilized to show the efficacy of the method for extracting a
gamma-ray signal from the background of hadronic generated images.
\end{abstract}
\begin{keyword}
Gamma-ray astronomy; Atmospheric Cherenkov Technique; Wavelets
\end{keyword}
\end{frontmatter}

\section{Introduction}
The study of astrophysical sources of very high energy (VHE, E $>$ 100
GeV) gamma rays was revolutionized by the development of the Imaging
Atmospheric Cherenkov technique (ACT), pioneered by the Whipple
Gamma-ray Collaboration in the late 1980s. The Collaboration utilizes
a 10-m optical reflector and camera consisting of an array of closely
packed photomultiplier tubes (PMTs) mounted in the focal plane. The
camera detects Cherenkov radiation produced by gamma-ray and
cosmic-ray air showers from which an image of the Cherenkov shower can
be reconstructed. The reflector, located at the Fred Lawrence Whipple
Observatory on Mt. Hopkins (elevation 2320 m) in southern Arizona, was
the first instrument to detect a VHE gamma-ray signal from the Crab
Nebula with high significance \cite{weekes89}. Since then numerous
modifications have been made to improve the sensitivity and
performance of the system which has been used as the model for many
additional observatories throughout the world.

The advancement of the ACT over the past decade has been directed
towards detecting lower energy gamma-rays, thereby closing the gap
between space-based instruments and ground-based observatories, and
improving the sensitivity to weaker gamma-ray sources. 
This has been accomplished by utilizing
finer pixellated cameras, faster electronics and more intelligent
triggering systems, for example the GRANITE III upgrade of the Whipple
Observatory's 10-m gamma-ray telescope \cite{bradbury99,lessard01}
and the Cherenkov at Th\'emis (CAT) telescope in the French Pyrenees
\cite{barrau98}. The push towards lower energy thresholds has
presented new challenges for the ACT. Firstly, the lower light levels
associated with low energy gamma-ray air showers resulted in
significant sensitivity to Cherenkov light from single local muons
\cite{catanese95}. Secondly, the small images recorded from low energy
showers cover only a few pixels making image reconstruction less
precise \cite{moriarty97}. Recent efforts in ACT have been directed
towards resolving these challenges.  We present in this paper a novel
method based on wavelets to enhance the image processing technique.

The paper is organised as follows. Section 2 provides an introduction to
the imaging atmospheric Cherenkov technique and the traditional
method used for image processing. The novel image
processing method based on wavelets is presented in Section 3. A comparison
with the traditional method is also discussed towards the end of that section. 
The effects of traditional cleaning versus 
wavelet cleaning on image reconstruction and characterization, are
presented in Section 4. Conclusions are included in Section 5.

\section{Imaging Atmospheric Cherenkov Technique}
Primary cosmic rays and gamma rays entering the atmosphere initiate
showers of secondary particles which propagate towards the
ground.  The trajectory of the shower continues along the path of the
primary particle.  If the optical reflector lies within the 300 m
diameter Cherenkov light pool, it forms an image in the PMT camera.
The appearance of this image depends upon a number of factors. The
nature and energy of the incident particle, the arrival direction and
the point of impact of the particle trajectory on the ground, all
determine the initial shape and orientation of the image. This image
is modified by the point spread function of the telescope, the
addition of instrumental noise in the PMTs and subsequent electronics,
the presence of bright stellar images in certain PMTs, the diffuse night
sky background and by spurious
signals from charged cosmic rays physically passing through the
tubes. Monte Carlo studies have shown that gamma-ray induced showers
give rise to more compact images than background hadronic showers and
are preferentially oriented towards the source position in the image
plane \cite{hillas85}. By making use of these differences, a gamma-ray
signal can be extracted from the large background of hadronic showers
and a gamma-ray map over the field of view (FOV) can be obtained. The
method of extracting a gamma-ray signal from the hadronic background
can be found in \cite{fegan94}.

\subsection{Traditional Image Processing}
\label{section:traditionalimageprocessing}
Prior to analysis of the recorded images, two calibration operations
must be performed: the subtraction of the pedestal analog-digital
conversion (ADC) values and the normalization of the PMT gains, a
process known as flat-fielding.

The pedestal of an ADC is the finite value which it outputs for an input without signal from genuine showers.
This is usually set at 20 digital counts so that small negative
fluctuations on the signal line, due to night sky noise variations,
will not generate negative values in the ADC. The pedestal for each
PMT is determined by artificially triggering the camera, thereby
capturing ADC values in the absence of genuine input signals. The PMT
pedestal and pedestal variance are calculated from the mean and
variance of the pulse-height spectrum generated from these injected
events.

The relative PMT gains are determined by recording a thousand images
using a fast Optitron Nitrogen Arc Lamp illuminating the focal plane
through a diffuser. These nitrogen pulser images are used to determine
the relative gains by comparing the relative mean signals seen by each
PMT.

\begin{figure}
\epsfig{file=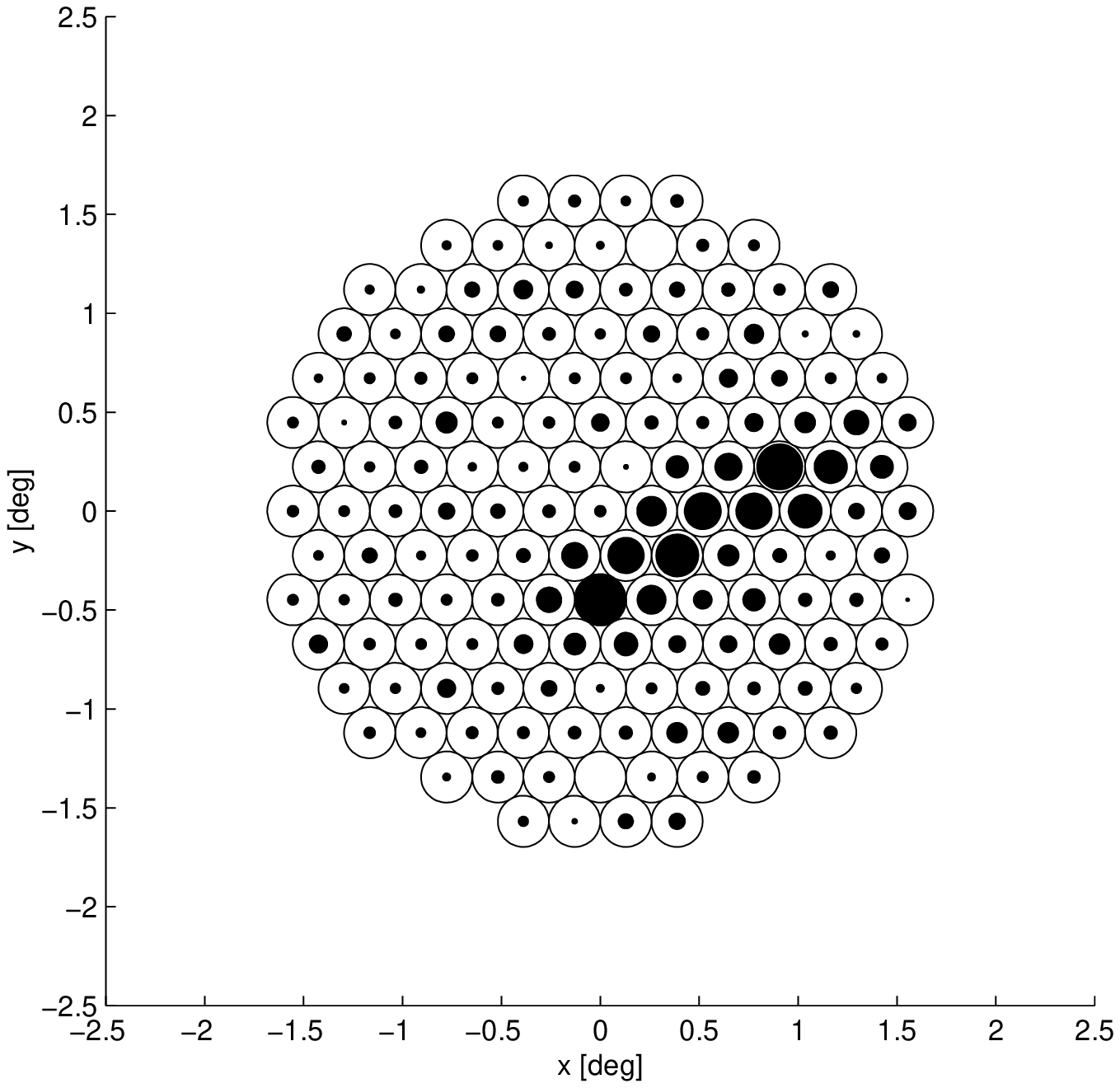,width=0.5\linewidth}
\epsfig{file=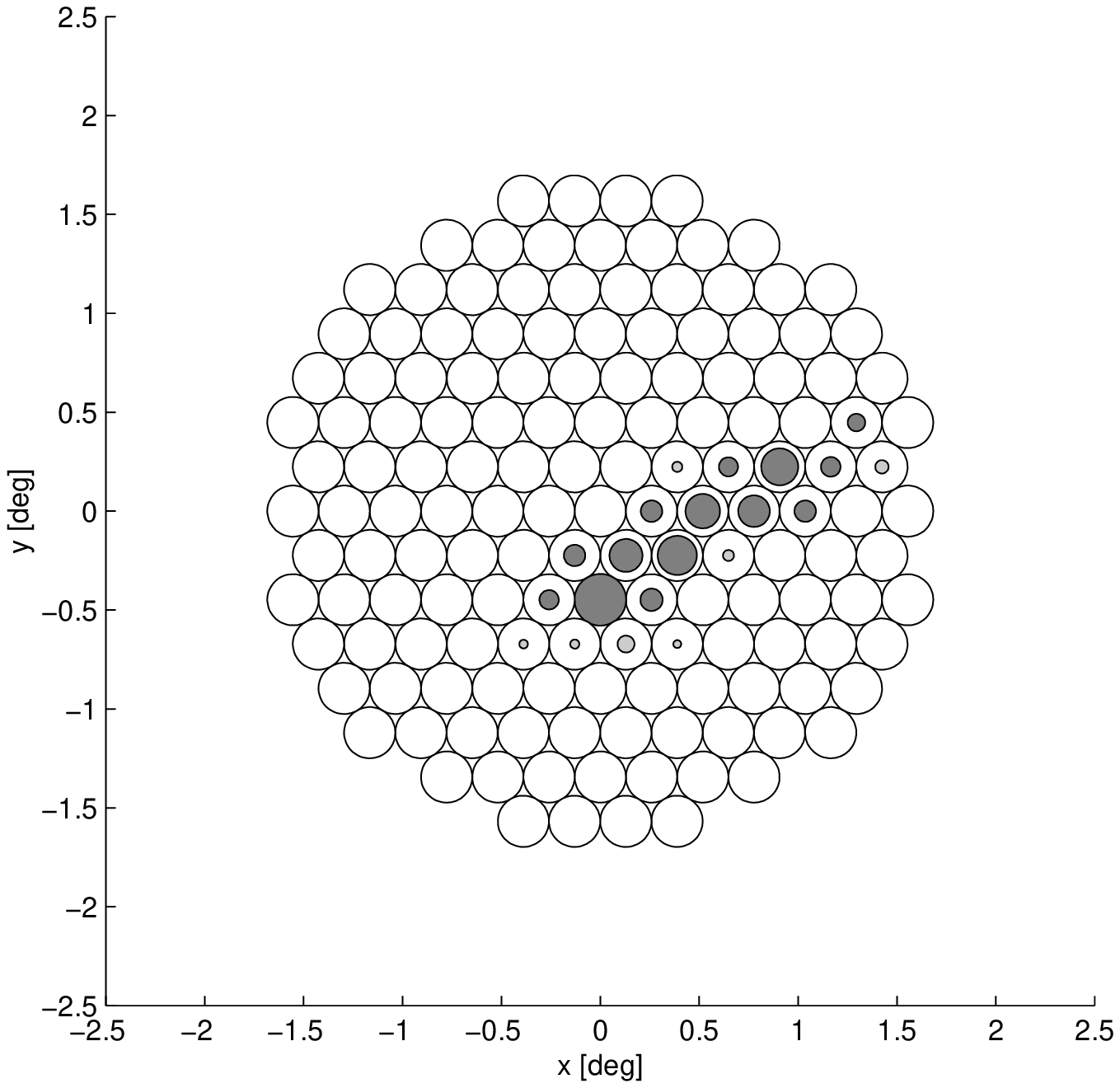,width=0.5\linewidth}
\caption{(Left) Example of an image prior to the processing procedure
given in the text. Diameter of each filled circle is proportional to
the ADC signal for that PMT. (Right) The same image after pedestal
subtraction, application of picture (dark gray filled circles) and
boundary (light gray filled circles) thresholds and gain
normalization.}
\label{figure:imclean}
\end{figure}

Fluctuations in the image usually arise from electronic noise and
night-sky background variations. The traditional method to reject
these distortions was developed for a camera consisting of 109 PMTs
(each viewing a circular field of 0.259$^\circ$ diameter) utilized by
the Whipple Observatory's 10-m gamma-ray telescope. The method selects
a PMT to be part of the image if it has a signal above a certain
threshold or is beside such a PMT and has a signal above a lower
threshold. These two thresholds are defined as the picture and
boundary thresholds, respectively. 
%The picture threshold is the
%multiple of the root mean square (RMS) pedestal deviation which a PMT's
%signal must exceed to be considered part of the picture. The boundary
%threshold is the multiple of the RMS pedestal deviation which PMTs
%adjacent to the picture must exceed to be part of the boundary. 
The picture and boundary thresholds are multiples of the root mean square
(RMS) pedestal deviation which PMT's signal must exceed to be considered
part of the picture or boundary, respectively.
The
picture and boundary PMTs together make up the image; all others are
zeroed. This image cleaning procedure is depicted in
Figure~\ref{figure:imclean}. A picture threshold of
$4.25\times$RMS and a boundary threshold of $2.25\times$RMS were
chosen to select the greatest number of PMTs with signal while at the
same time limiting the inclusion of PMTs with noise alone. 

Monte Carlo simulations of gamma-ray induced air showers are used to
test the performance of these imaging procedures. The simulations
track each particle produced in the air shower and trace each
Cherenkov photon emitted to the image plane. Included in the
simulation are the effects of the atmosphere, mirror alignment and
reflectivity, and the quantum efficiency of the PMTs. Background light
from the night sky and electronic noise are added to match the
conditions present during the course of typical observations. These
simulations, described in \cite{kertzman94}, are utilized by the
Collaboration to determine the energy dependent collection area of the
Whipple 10-m telescope as presented in \cite{mohanty98}.

\begin{figure}
\epsfig{file=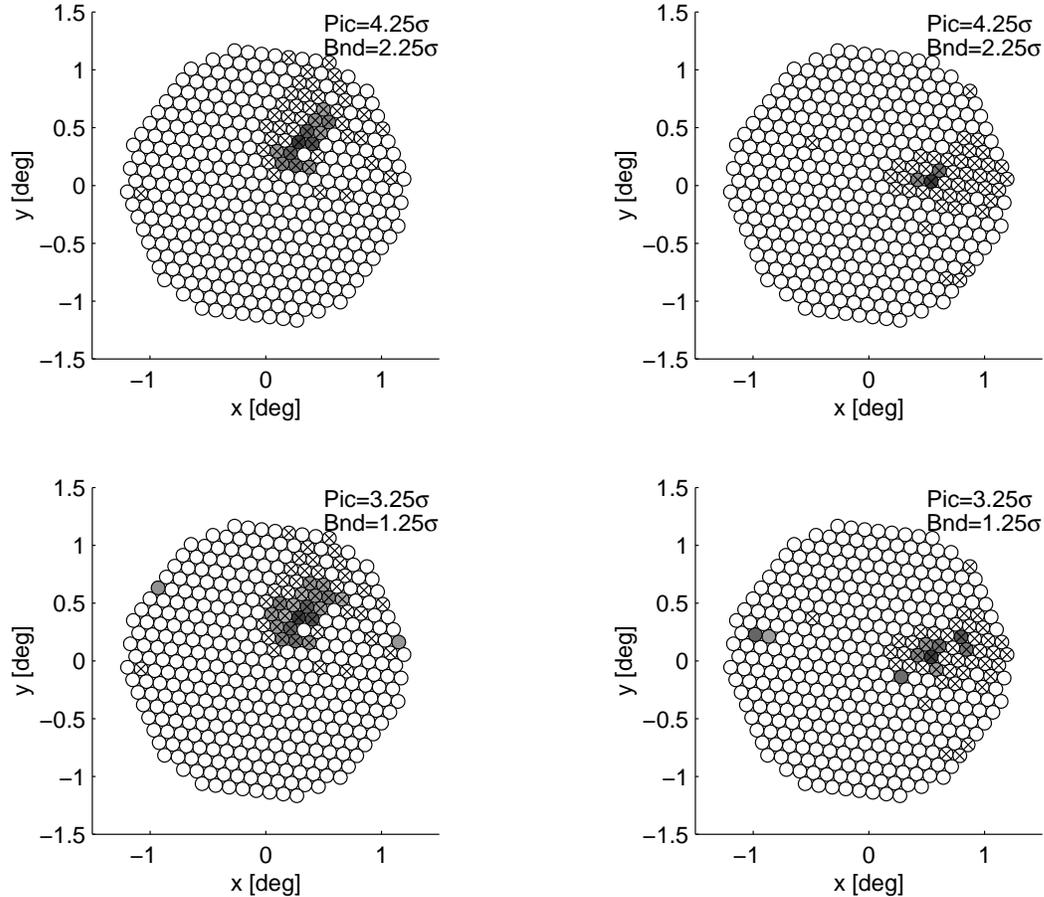,width=\linewidth}
\caption{(Top) Two examples of images cleaned using the traditional
method of picture and boundary thresholds given in the text. Pixels
selected by this technique are shaded in gray scale with darker shades
corresponding to greater ADC signal for that PMT. The x symbols
indicate pixels with at least one photoelectron due to Cherenkov
light. (Bottom) The same images but cleaned using lower picture and
boundary thresholds.}
\label{figure:sample_images_picbnd}
\end{figure}

\begin{figure}
\epsfig{file=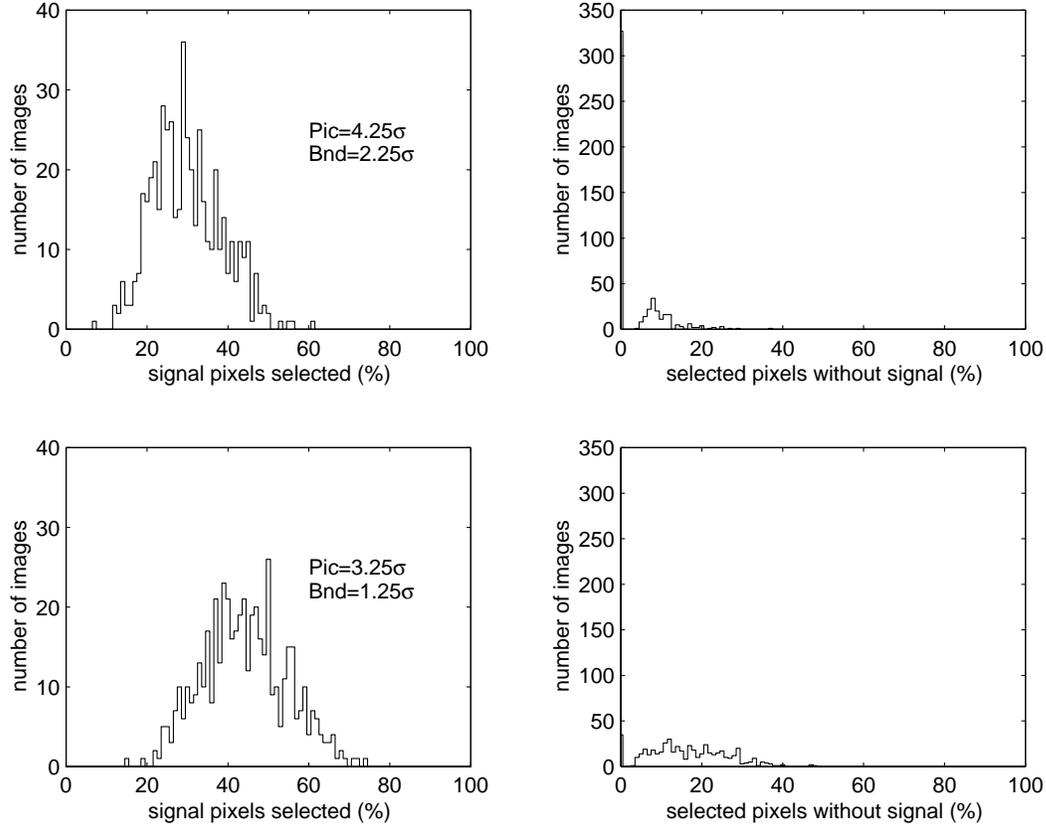,width=\linewidth}
\caption{(Top) Distribution of the number of (Left) signal pixels
selected which have at least one Cherenkov photon, relative 
to the total number selected. The right panel shows the number of pixels selected without signal,
i.e. noise (zero Cherenkov photons), relative to the total number selected. In 
both panels, the traditional method of picture boundary thresholds 
given in the text was used. (Bottom) The same distributions but utilizing
lower picture boundary thresholds.}
\label{figure:selected_pixels_picbnd}
\end{figure}

Figure~\ref{figure:sample_images_picbnd} shows two sample images of
Monte Carlo simulated showers initiated by 150 GeV gamma-rays.  The x
symbols indicate PMTs with at least one photoelectron due to Cherenkov
light from the shower. The pixels selected by the traditional picture
and boundary threshold method are shown in gray scale with darker
shades corresponding to greater ADC signal for that PMT. For the image
on the top left, 25\% of the PMTs with Cherenkov signal are selected by
the cleaning process. The image on the top right was chosen to show an
example where very few pixels are selected by the standard cleaning
method. In this case just three pixels or 7\% of the PMTs with signal
were selected. By lowering the picture and boundary thresholds we can
accept more signal PMTs, as shown in the bottom panels of
Figure~\ref{figure:sample_images_picbnd} which depict the same images
given in the top panels but cleaned with picture and boundary
thresholds of $3.25\times$RMS and $1.25\times$RMS respectively.
However, the number of noise PMTs selected with the lower thresholds
also increases. The goal of the image cleaning process is to select
all of the PMTs with signal while rejecting all PMTs with noise
alone. We have simulated 500 images (incident energy 143 GeV) 
and determined the number of
signal PMTs correctly selected and the number of noise PMTs
incorrectly selected. These results are shown in
Figure~\ref{figure:selected_pixels_picbnd}.

High energy gamma-ray showers produce large, bright images which
result in a greater number of PMTs being selected. This is adequate
for the characterization of the image shape and orientation used for
discriminating gamma-rays from the more numerous hadrons. However, for
low energy showers, only a few PMTs are selected by the traditional
method resulting in a poor reconstruction of the image. Relaxing the
thresholds includes too many noise PMTs, as shown in
Figure~\ref{figure:selected_pixels_picbnd}, resulting in increased
error in the reconstruction which affects the sensitivity, energy and
angular resolution of ACT telescopes operating at low energies.

\section{Wavelet Method}
Wavelet techniques are developing rapidly and are proving to be very
efficient as signal and image processing methods. These techniques
have been applied in different fields and in particular in
astrophysics and cosmology. Among the myriad of papers, examples of
the detection of structure in astrophysical and cosmological images
using wavelets can be found in 
\cite{damiani97,krywult99,lazzati99,cayon2000}. De-noising and
compression of astrophysical and cosmological images have also been
performed using wavelet techniques \cite{sanz99a,sanz99b,tenorio99}.
Wavelets are also a useful tool for performing statistical analysis,
for example see \cite{hobson99,barreiro2000}.  As
previously described, the cleaning and characterization of Cherenkov
images are not easy tasks and have to be performed by applying well
adapted methods.  Wavelets are just now being utilized to analyse
these images. Wavelet moments were introduced as a complementary
method to characterize Cherenkov images in
\cite{haungs99}. 
%Before any characterization is done one has to
%clean the Cherenkov images. Selection of signal pixels should be done
%so that the final cleaned image contains as many pixels with Cherenkov
%photons as possible while 
%nly a few noise pixels are included.
%including only a few noise pixels. 
Wavelets could as well be used to clean and characterise Cherenkov images at
the same time. In this work, we will use wavelets, in a 
novel way,  only to select 
pixels containing Cherenkov photons. That is, wavelets will be 
used for the cleaning process. A Hillas parametrisation will afterwards
provide the characterisation of the images.

\subsection{Image Processing Using Wavelets}
We have developed a wavelet based method to select signal pixels in
atmospheric Cherenkov images. Wavelet decomposition of an image
provides information about the contribution of different scales to
each pixel. The combined information at several scales at each
location in the image makes wavelet methods a very powerful technique. 
Wavelet
coefficients $wv(R,\vec b)$ are calculated for a fixed scale $R$ at
each pixel $\vec b$, by convolving the image under analysis $f(\vec
x)$ with a wavelet $\Phi=\Phi(\vec x, R)$:
\begin{equation}
wv(R,\vec b)=\int d\vec x f(\vec x)
\Phi \bigl( {{\vert \vec x-\vec b\vert }\over {R}}\bigr).
\end{equation}
The method we use to separate the noise from the signal relies
on the characteristics of the noise rather than on the characteristics
of the signal. These characteristics will be determined by the
distributions of wavelet coefficients corresponding to noise, at
several scales. As one of the possible analyzing wavelets, we have
chosen the so called Mexican Hat wavelet given by
\begin{equation}
\Phi \bigl( {{\vert \vec x-\vec b\vert}\over {R}} \bigr) =
\bigl( 2-{{\vert\vec x-\vec b\vert ^2}\over {R^2}}\bigr)
e^{-{{\vert\vec x-\vec b\vert ^2}\over {2R^2}}} 
{{1}\over {(2\pi)^{0.5}R}}.
\end{equation}
An isotropic wavelet such as the Mexican Hat seems more appropriate for
this analysis than an anisotropic one, since we do not want any direction
to be preferred by the wavelet coefficients.

%\begin{figure}
%\epsfig{file=mexhat.eps,width=\linewidth}
%\caption{The Mexican Hat wavelet.}
%\label{figure:mexicanhat}
%\end{figure}

As previously mentioned, the method designed to select signal pixels
in our images takes into account the {\it a priori} information about the
characteristics of the noise. The pedestal and pedestal variance for
each pixel are determined as described in
$\S$\ref{section:traditionalimageprocessing}. The noise at each
location is approximately described by a Gaussian distribution with
pedestal mean, and variance equal to the pedestal variance. 
A characterisation of the noise distribution affecting Cherenkov
images was presented in figure 3 in \cite{fegan94}. As one can see 
the fit to a Gaussian distribution is very good.
We
generate 300 noise simulations (determined to be a sufficient number of
simulations to reproduce the distributions; this number is a 
good compromise between accuracy and computational speed) 
and for each simulation we
calculate the wavelet coefficients corresponding to four different
scales, multiples of the characteristic pixel scale $R=2\times
pix\_size,3\times pix\_size,4\times pix\_size,5\times pix\_size$. At
each pixel and each scale we generate the wavelet coefficient
probability distributions corresponding to noise.

Once the noise wavelet coefficient distributions are known at each
location of the image for the four scales
%, we proceed as follows.  We
%first calculate 
the wavelet coefficients corresponding to the image
under analysis are calculated for the same four scales. 
In the second step 
%consists in assigning 
a probability is assigned to each of these wavelet coefficients by
comparing their values with the corresponding noise wavelet
coefficient distribution. Finally only those pixels with
wavelet coefficients outside the noise wavelet coefficient distributions 
at all scales are selected. The use of several scales in the wavelet method has the
advantage of using information about the noise not only in the pixel
under analysis (as the traditional picture-boundary selection method)
but in the neighbouring pixels. Note, that this is not the way in which
wavelets are normally used to separate signal from noise. In a 
``classical'' approach wavelet coefficients at scales dominated 
by noise (the smaller ones) are thresholded (hard or soft thresholding). 
The image is afterwards reconstructed based on these ``modified'' wavelet
coefficients. This is in brief, the ``classical'' denoising procedure
using wavelets. We would like to remark that we are not denoising the
images. Wavelets are here used to select pixels containing Cherenkov photons.
Noise will still be present in the image afterwards characterised with
Hillas parameters.

\begin{figure}
\epsfig{file=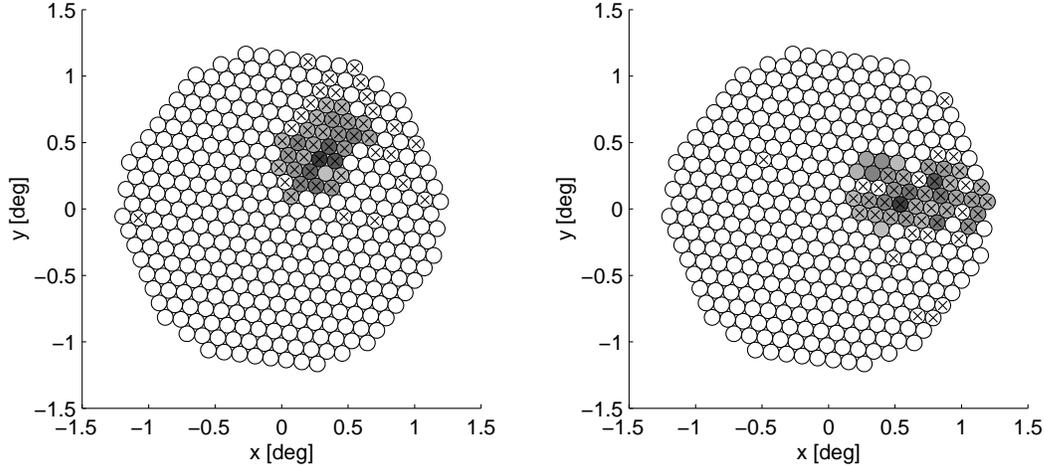,width=\linewidth}
\caption{The same sample images shown in
Figure~\ref{figure:sample_images_picbnd} but cleaned using the wavelet
method given in the text. Pixels selected by this technique are shaded
in gray scale with darker shades corresponding to greater ADC signal
for that PMT. The cross symbols denote pixels with at least one
photoelectron due to Cherenkov light.}
\label{figure:sample_images_wv}
\end{figure}

\subsection{Results}
\begin{figure}
\epsfig{file=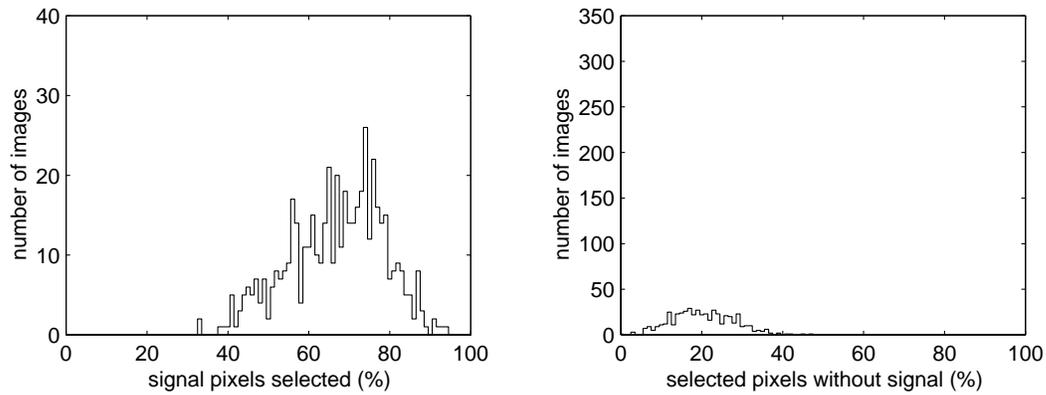,width=\linewidth}
\caption{Distribution of (Left) the number of selected signal pixels relative to the 
total number of selected pixels
and (Right) the number of selected pixels without signal, i.e. noise, 
relative
to the total number of pixels selected,
using the wavelet method given in the text.}
\label{figure:selected_pixels_wv}
\end{figure}

The same Monte Carlo simulations used to test the performance of the
traditional cleaning procedure are now utilized to determine the
characteristics of the wavelet method.
Figure~\ref{figure:sample_images_wv} shows the same images depicted in
Figure~\ref{figure:sample_images_picbnd}, but cleaned with the wavelet
method. This procedure selected significantly more signal pixels than
the picture boundary method, while at the same time, selecting only a
few pixels which do not contain signal due to Cherenkov light from the
air shower. For the image on the left, 57\% of the signal pixels are
selected, compared to the 25\% selected by the traditional method, as
shown in Figure~\ref{figure:sample_images_picbnd}.  For the image on
the right 65\% of the signal pixels are selected compared to the 7\%
selected by the traditional method, as shown in
Figure~\ref{figure:sample_images_picbnd}. The distribution of the
number of signal pixels selected and the number of noise pixels
erroneously chosen for 500 simulated gamma-ray images are depicted in
Figure~\ref{figure:selected_pixels_wv}. Clearly, the wavelet method
includes more of the Cherenkov light signal than the traditional
method of picture boundary thresholds. This results in a greater
number of available pixels to reconstruct the characteristics of the
Cherenkov light image. 
The pixels selected by this method and not by the traditional method must
have small signal and thus greater noise contribution.  The pulse
height distribution for a single pixel is represented in Figure 6. The
pulse height is the sum of signal and noise contributing to a
pixel. The noise is approximately Gaussian in shape. The signal lies
under the noise distribution and is responsible for the tail at larger
pulse heights. As one can see in the top panel, the picture boundary
thresholds indiscriminately discard pulses below the fixed threshold,
some with genuine signal, albeit small. On the other hand, the wavelet 
cleaning method selects more pulses with a few well inside the noise
distribution. This is a result of the wavelet method including
more information about the spatial distribution of the noise, and not
just the noise itself, thereby allowing the inclusion of pixels with
low signal to noise contributions. It is this property that makes the
wavelet method optimal for subsequent image reconstruction and
characterization based on Hillas parameters.

\begin{figure}
\epsfig{file=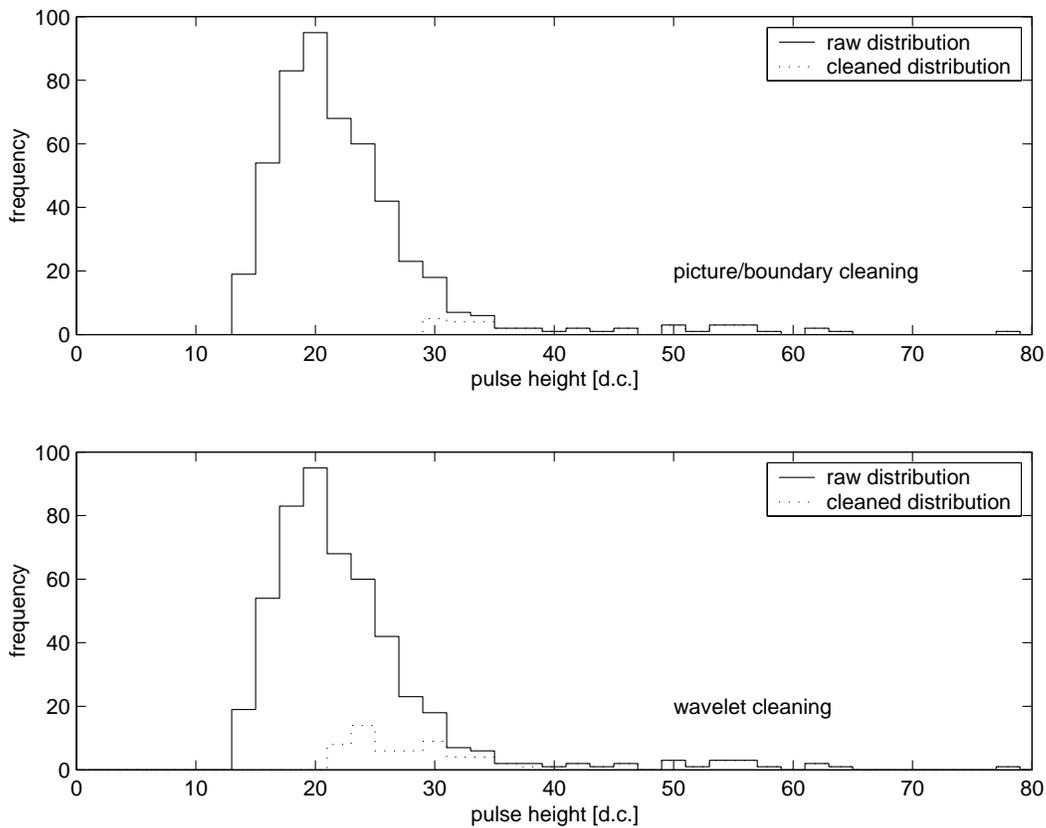,width=\linewidth}
\caption{Distribution of pulse height of a pixel before (solid line) 
and after (dotted line) applying a cleaning method. 
The result of applying the picture/boundary method is presented in the top panel. Wavelet cleaning results in the pulse height distribution shown
in the bottom panel.}
\label{figure:pulse_height}
\end{figure}

\section{Image Reconstruction and Characterization}
\begin{figure}
\epsfig{file=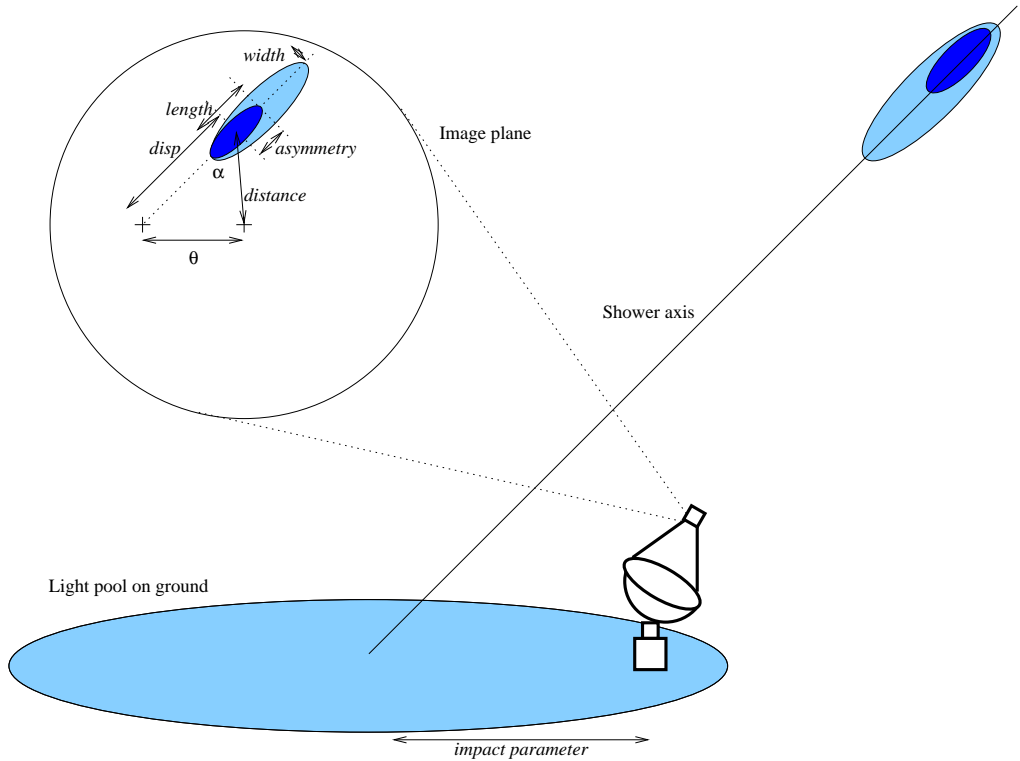,width=\linewidth}
\caption{Depiction of the light produced by air showers. The image
plane shows the definition of the Hillas parameters used to
characterize each image.}
\label{figure:hillasparameters}
\end{figure}

Each Cherenkov image is characteried using a moment analysis
\cite{reynolds93}. The roughly elliptical shape of the image is
described by the length and width parameters. Its location
and orientation within the FOV are given by the distance and
$\alpha$ parameters, respectively. The asymmetry parameter,
defined as the third moment of the light distribution, describes the
skew of the image along its major axis. Also determined are the two
highest signals recorded by the PMTs (max1, max2) and the amount
of light in the image (size). These parameters are defined in
Table~\ref{table:hillasparameters} and are shown in
Figure~\ref{figure:hillasparameters}. These are called Hillas parameters
\cite{hillas85}.

\begin{table}
\caption{Definition of image parameters, used to
characterize the image shape and orientation in the FOV (see
Figure~\protect\ref{figure:hillasparameters}).\label{table:hillasparameters}}
\vspace{0.5cm}
\begin{tabular}{ll}\hline
Parameter & Definition                                                \\ 
\hline
max1:     & largest signal recorded by the PMTs.                      \\
max2:     & second largest signal recorded by the PMTs.               \\
size:     & sum of all signals recorded.                              \\
centroid: & weighted center of the light distribution ($x_c,y_c$).\\
width:    & the RMS spread of light along the minor axis of the image;\\
          & a measure of the lateral development of the shower.       \\
length:   & the RMS spread of light along the major axis of the image;\\
          & a measure of the vertical development of the shower.      \\
distance: & the distance from the centroid of the image to the center \\ 
          & of the FOV.                                               \\
$\alpha$: & the angle between the major axis of the image and a line  \\
          & joining the centroid of the image to the center of the    \\
          & FOV.                                                      \\
asymmetry:& the skewness of the light distribution relative to the    \\
          & image centroid.                                           \\ 
\hline
\end{tabular}
\end{table}

Gamma-ray events give rise to shower images which are preferentially
oriented towards the source position in the image plane. These images
are narrow and compact in shape, elongating as the impact parameter
increases. They generally have a cometary shape with their light
distribution skewed towards their source location in the image
plane. Hadronic events give rise to images that are, on average,
broader (due to the emission angles of pions in nucleon collisions
spreading the shower), and longer (since the nucleon component of the
shower penetrates deeper into the atmosphere) and are randomly
oriented within the FOV. Utilizing these differences, a gamma-ray
signal can be extracted from the large background of hadronic showers.

\begin{figure}
\epsfig{file=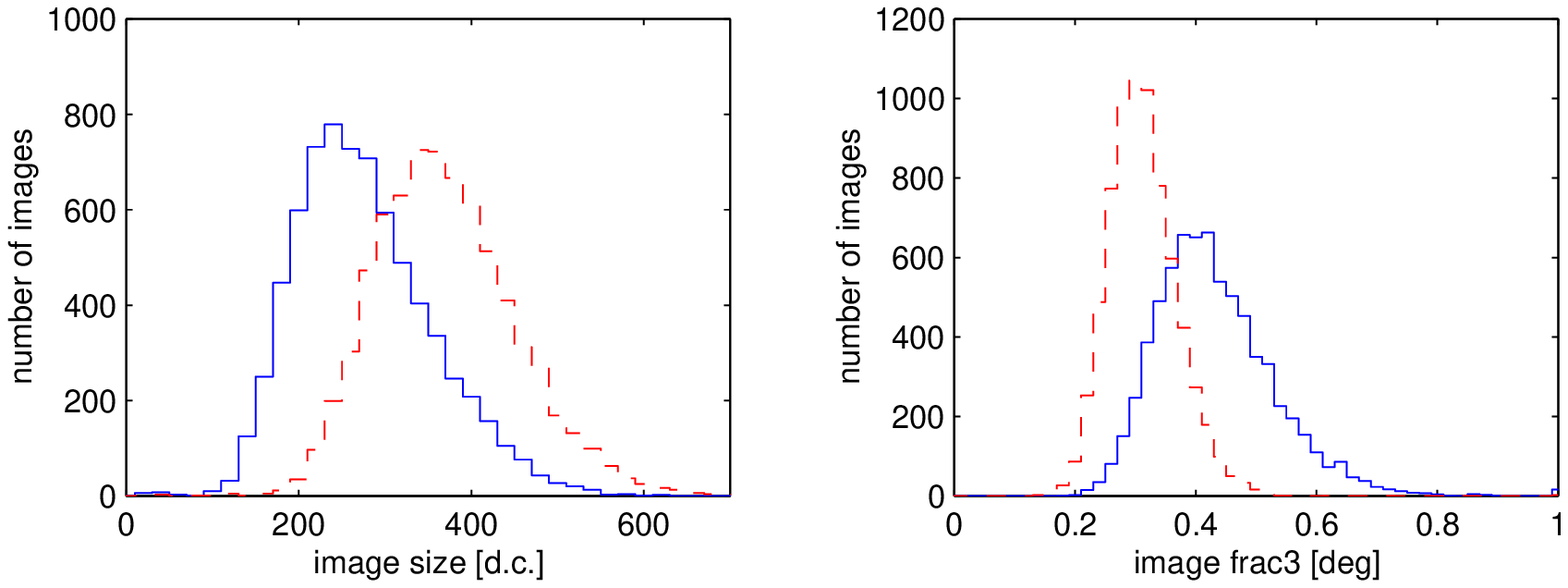,width=\linewidth}
\epsfig{file=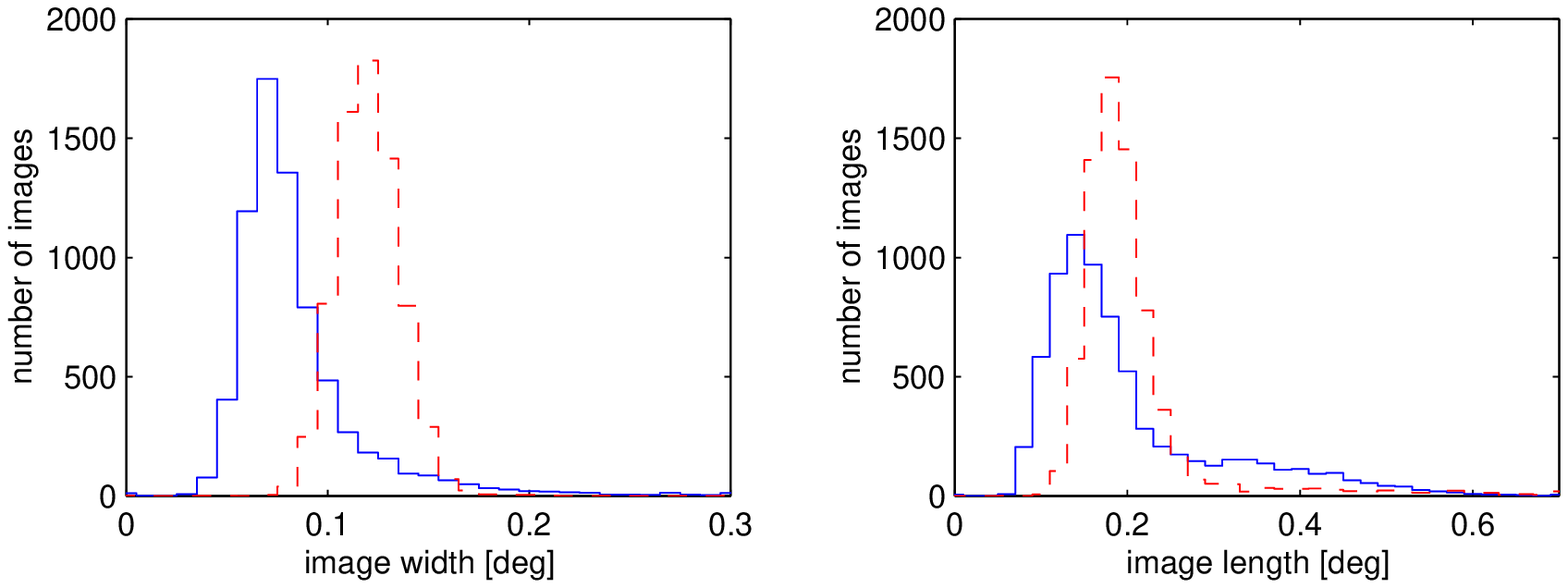,width=\linewidth}
\epsfig{file=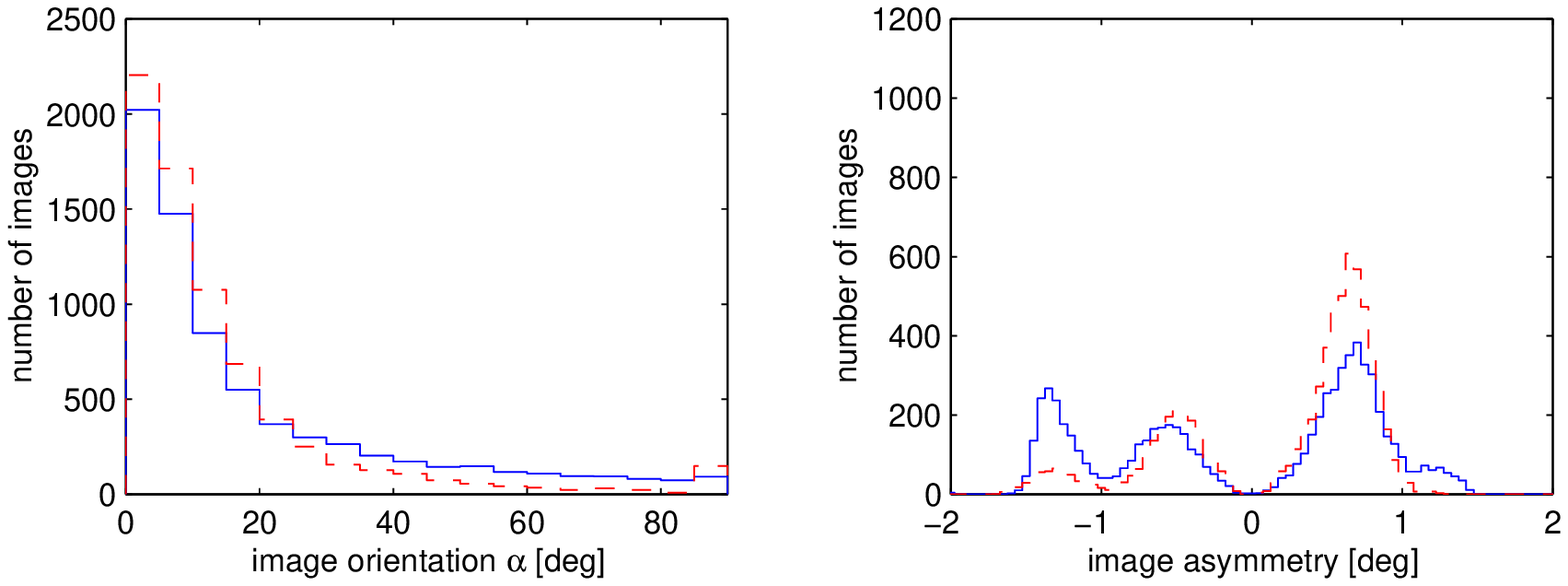,width=\linewidth}
\caption{Distribution of the Hillas parameters for 7156 Monte Carlo
simulations of 150 GeV gamma-ray induced air showers. Results from the
traditional picture-boundary cleaning method are shown as solid
lines. Results from the wavelet cleaning method are shown as dashed
lines.}
\label{figure:parameter-distributions}
\end{figure}

Monte Carlo simulations of gamma-ray induced air showers are used to
visualize the effect of wavelet cleaning on the moments of the light
distribution. Firstly, as depicted in
Figure~\ref{figure:parameter-distributions}, the wavelet cleaned
images tend to have greater size and a lower concentration of light in
the brightest three pixels (called frac3). These are simply due to the
inclusion of a greater number of signal pixels in the image. Secondly,
for the same reason, the images tend to be wider and longer as shown
in Figure~\ref{figure:parameter-distributions}. Also noted in this
figure is that the width and length distributions for the wavelet cleaned
images are more symmetric and lack the long tails at large values of
width and length as found with the picture-boundary cleaned
images. These tails are likely the result of ``hot'' single pixels
which grossly distort the image. Lastly, images are selected as
gamma-ray candidates if they have small $\alpha$ values and positive
asymmetry. The greater number of pixels selected by the wavelet
cleaning method enables a more accurate determination of both of these
quantities as shown in Figure~\ref{figure:parameter-distributions}.
When extracting a gamma-ray signal from a background of hadronic
images, a selection of $\alpha < 15^\circ$ and asymmetry $>$ 0 is
made. In the case of the $\alpha$ selection, the wavelet method
collects 15\% more images.  When combined with the asymmetry selection
33\% more images are collected due to the improved reconstruction.

\subsection{Application to observations taken on established sources}

   The efficacy of utilizing wavelets as a method for image processing for
   imaging Atmospheric Cherenkov telescopes can be verified by applying the
   technique to observations of established sources such as the Crab Nebula,
   the standard candle for ground-based gamma-ray astronomy. We compare our
   results to traditional analysis methods used by the Whipple Collaboration
   and outlined in \cite{fegan94}. The traditional analysis 
   technique consists of the
   image processing methods described in this paper, including the application
   of picture and boundary thresholds for image cleaning, followed by
   characterization and image selection criteria known as Supercuts \cite{reynolds93}. These
   selection criteria, preferentially select gamma-ray induced images from the
   much more numerous hadronic images. Our comparison follows the same
   techniques however, the images will be cleaned with the wavelet method
   described herein. 

\begin{figure}
\epsfig{file=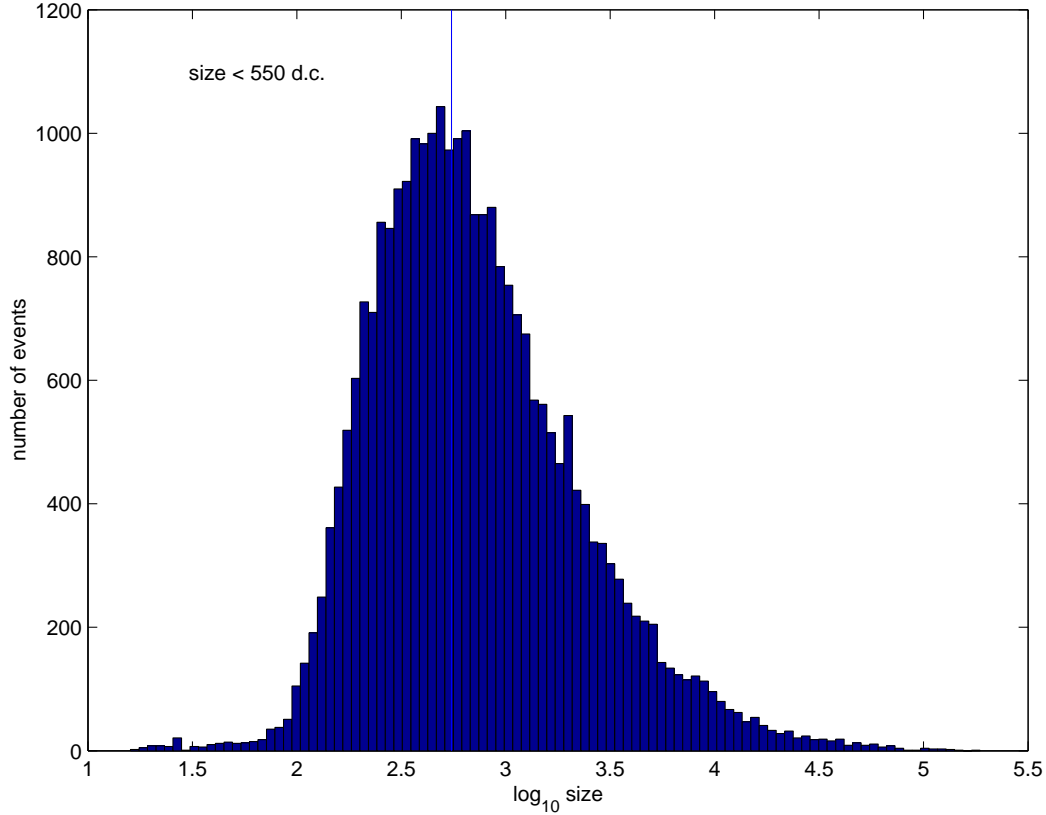,width=\linewidth}
\caption{The distribution of image size as derived from images cleaned with the
   traditional method of picture boundary thresholds. The vertical line
   indicates the upper bound in size chosen for the analysis.}
\label{figure:SizeDistribution}
\end{figure}

   As previously described, wavelet cleaning is expected to provide improved
   performance for low energy events, where the number of signal pixels
   selected by traditional methods is small and thus prone to greater errors in
   the reconstruction. In fact, Supercuts are optimized for peak sensitivity
   discarding many low energy images, as they cannot be distinguished from
   single muon images and noise. As shown in Figure~\ref{figure:SizeDistribution}., the peak in the
   distribution of image size selected by Supercuts is 550 d.c., accordingly
   for comparison, we choose to focus on images below this size.

%   [ I need to do this figure still].
%   Figure 8.

   Our comparison requires two distinct procedures. Firstly, image selection
   cuts must be optimized for both the picture boundary and
   wavelet cleaning methods. Note that cuts for the picture boundary cleaning
   method are re-optimized from traditional values due to the introduction of a
   size limitation which alters the optimum cuts, see \cite{moriarty97}
   for an in depth discussion. Secondly, these new criteria must be
   tested on an independent data set to alleviate any bias due to optimization.
   In addition, for this analysis, we determine image size using a common
   definition to ensure that the same events are included in both cases; for
   these results image size was calculated from the picture boundary cleaned
   image. 

\begin{figure}
\epsfig{file=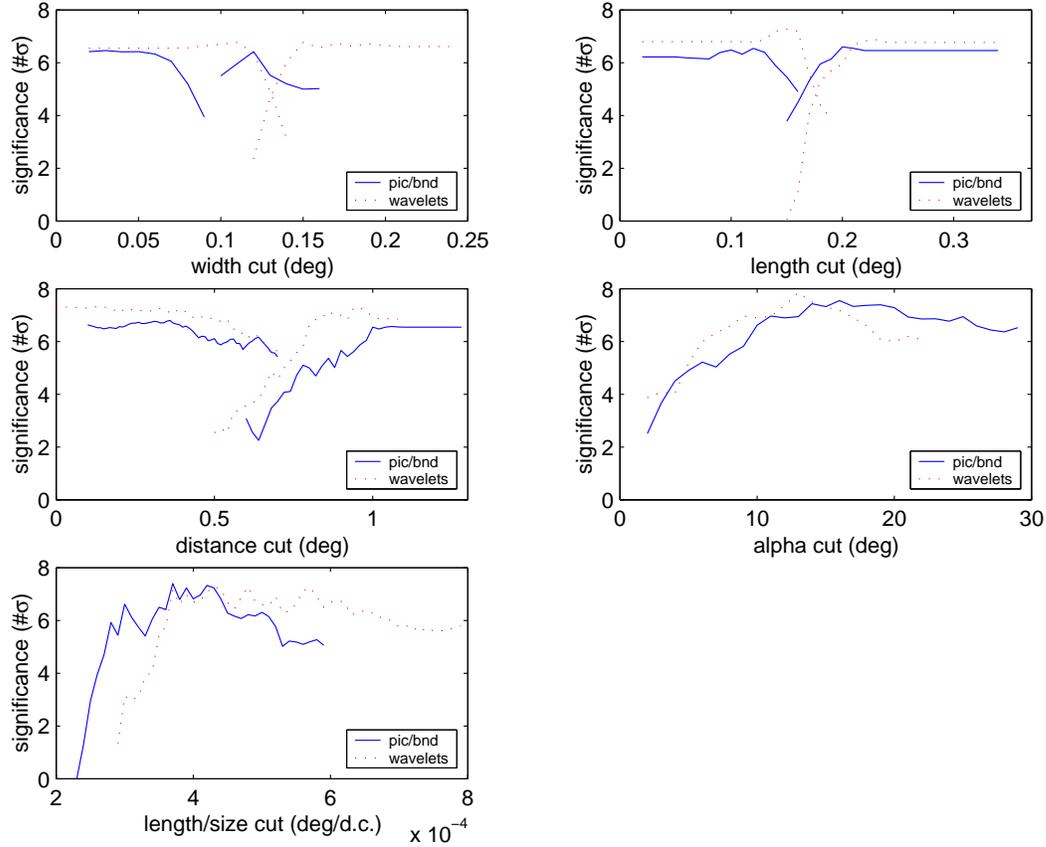,width=\linewidth}
\caption{Optimization of image selection cuts. The results for the data cleaned with
   the traditional picture boundary thresholds are shown as solid curves. The
   results for the data cleaned with wavelets are shown as dashed curves.
The two independent branches appearing in the width, length and distance plots
correspond to the optimization of the lower and upper limits.}
\label{figure:optimization}
\end{figure}

   The optimization procedure involves bounding all but one selection criterion
   whilst searching for optimum signal to noise performance via standard grid
   search methods. We define signal to noise performance as the number of
   standard deviations the signal appears above background. We have chosen a
   set of observations on the Crab Nebula, taken under good sky conditions and
   instrument operations. A total of 12 hours of data, taken over the period
   December 1999 to January 2000, are included. The results are shown in Figure~\ref{figure:optimization} and given in Table 2. The order that each parameter is optimized is as
   shown in Figure~\ref{figure:optimization}, starting from the top and working left to right. Before
   each parameter is optimized the traditional cut parameters were chosen, as
   given in Table 2. Upon the determination of an optimum cut, this new value
   is used for the remaining procedure. This method has been the standard
   practice of optimization by the Whipple Collaboration over the past decade
   and has been found to be more effective than other optimization methods such
   as the Simplex algorithm.

\begin{table}
\caption{Supercuts selection criteria, which were optimized on
contemporaneous Crab Nebula data.}
\vspace{0.5cm}
\begin{tabular}{lll}\hline
Traditional & Low Energy Images & Wavelet Cleaned\\
supercuts 2000 & Pict/Bound Cleaned & \\
\hline
   size $>$ 0 & 0 $<$ size $<$ 550 & 0 $<$ size $<$ 550                      \\
   max2 $>$ 30 &      max2 $>$ 30 &     max2 $>$ 30                      \\
   0.05 $<$ width $<$ 0.12  &  0.03 $<$ width $<$ 0.12  &  0.11 $<$ width $<$ 0.15 \\
   0.12 $<$ length $<$ 0.25 &  0.12 $<$ length $<$ 0.21 &  0.15 $<$ length $<$ 0.23 \\
   0.40 $<$ distance $<$ 1.00 & 0.36 $<$ distance $<$ 1.00 & 0.00 $<$ distance $<$ 0.94 \\
   $\alpha < 15$   &  $\alpha < 16$  &  $\alpha < 13$ \\
   length/size $<$ 0.00040 & length/size $<$ 0.00042 & length/size $<$ 0.00057\\
\hline
\end{tabular}
\end{table}

   The results of the optimization indicate enhanced reconstruction
   capabilities as shown by the inclusion of events at all lower distances and
   a lower cut of the image pointing angle alpha. Supercuts typically discard
   images at lower distance due to the potential for greater error in the
   calculation of the image pointing angle alpha.

   These results cannot be used to compare performance due to the optimization
   bias. As a result we have analyzed a set of observations taken on the Active
   Galactic Nuclei Markarian 421 and 501, long established as
   gamma-ray sources
   \cite{punch92,quinn96}. A total of 5 hours of data, taken
   between January 2000 and May 2000, under good sky and instrument operation,
   are included (data from Markarian 421 and 501 are gathered together to 
   provide a good unbiased source of VHE photons). The data used
   spanned the same range of elevation as the Crab data. All sets
   of data were taken at small zenith angles (zenith angle
   dependence is too small to measure). 
   The results, given in Table 3, show the expected
   increase in collection area for low energy events, afforded by the wavelet
   method.

\begin{table}
\caption{Results of the analysis of data taken on the Active Galactic
   Nuclei, Markarian 421 and 501. The increase in gamma-ray rate is the result 
of larger collection area at low energies afforded by the 
wavelet technique. In the traditional method these photons would have been
discarded due to lack of signal after cleaning thus lowering collection area /
efficiency.}
%Due to the highly variable emission states of
%   these objects the data selected showed a greater than 2 gammas/min flux as
%   derived from standard Whipple Observatory quicklook results.}
\vspace{0.5cm}
\begin{tabular}{lllll}\hline
   Image Cleaning &  ON source & OFF s.c. &
   Excess ($\#$ of $\sigma $)&  Rate (/min) \\
   Method & counts & & & \\
\hline
   Picture/Boundary   &     1446  &  1219 &   4.39  &  0.74 +/-0.17 \\
   Wavelets    &    2254 &   1902  &  5.46  &  1.02 +/- 0.21\\
\hline
\end{tabular}
\end{table}

%EFFICACY OF TECHNIQUE TESTED USING DATA TAKEN ON ESTABLISHED SOURCES
%OF TEV GAMMA-RAYS

%0. FOCUS ON LOW ENERGY EVENTS
%   o STANDARD SUPERCUTS ARE OPTIMIZED FOR PEAK SENSITIVITY. IN DOING 
%     SO MANY LOW ENERGY EVENTS ARE DISCARDED BECAUSE THEY CANNOT BE 
%     DISTINGUISHED FROM SINGLE MUON IMAGES. IF A CUT ON IMAGE SIZE IS
%     OPTIMIZED FOR SENSITIVITY TYPICALLY SIZES LESS THAN 550 D.C ARE 
%     DISCARDED. THE STANDARD PRACTICE HOWEVER IS TO UTILIZE A LENGTH
%     DEPENDENT SIZE CUT WHICH IS SLIGHTLY BETTER FOR RETAINING IMAGES
%     WITH SIZE GREATER THAN 350 D.C. WE WILL THUS FOCUS ON LOW ENERGY
%     EVENTS WITH SIZE < 550 D.C.

%(plot) optimum size cut

%1. DETERMINE OPTIMAL CUTS.

%(plot) optimization pic-bnd and wavelets

%   o IMPROVED ALPHA CUT DUE TO BETTER RECONSTRUCTION
%   o APPARENT GREATER COLLECTION AREA - BUT WILL RESERVE CONCLUSION UNTIL
%     APPLIED TO INDEPENDENT DATA SET

%2. APPLY OPTIMUM CUTS TO INDEPENDENT DATA SET. VOILA.

%3. COMBINED WITH CUTS ABOVE 550 ?

\section{Conclusions}

The imaging atmospheric Cherenkov technique has pioneered the
   detection of VHE gamma-rays from ground-based observatories. The main
   difference remaining between space-based and ground-based gamma-ray
   observatories is the detection of gamma rays in the 10 to 200 GeV
   energy band. Photons in this energy range are too few for the limited
   collection areas of space-based instruments and typically produce low
   levels of Cherenkov light for a single reflector at ground level.
The aim of this paper
was to present a novel cleaning method based on wavelets 
that increases the number of signal pixels selected while rejecting 
the maximum number of noise pixels. 

The traditional cleaning method used to analyse images provided
by Cherenkov telescopes selects those pixels which are
above the picture threshold (4.25xRMS) or are beside 
such pixels and have signal above the lower neighbour threshold
(2.25xRMS). 
The performance of a cleaning method can be viewed using Monte Carlo 
simulations. Figures 2 and 3 show these results in the case of the 
traditional image cleaning method. Only a small percentage (average of
$\sim 30\%$) of the
total number of PMTs that have real signal are selected by this method. 
The percentage can be increased by lowering 
the picture and boundary thresholds. However, this also
increases the number of noise pixels incorrectly selected as signal 
pixels (see bottom panels in Figures 2 and 3).

A method to select a greater number of real signal pixels, while
excluding pixels with noise alone has been presented in this work. 
The method is based on the significant property of wavelets of providing
information about the contribution of different scales to each 
location of an image. This is a novel method and it 
is important to notice that denoising is not carried out.    
Convolution of an image with a wavelet
of certain size will provide a new image (in wavelet space) with
information at each pixel of the contribution of the scale given by the
wavelet size. 
The Mexican Hat wavelet was chosen for this work based on its isotropy. 
Assuming the noise in each pixel comes from
a Gaussian distribution of well characterized mean and variance, we build the
probability distribution of noise wavelet coefficients at four
scales. Comparison of the wavelet coefficients obtained for the image
at the four analysed scales with the noise wavelet coefficient 
distributions allows us to discriminate between signal and noise dominated
pixels. As shown in Figures 4 and 5 the percentage of signal 
selected pixels (average of $\sim 70\%$) 
has increased compared to
the traditional cleaning method. At the same time the number of
erroneously selected pixels has only slightly increased.

The fact that the wavelet method selects a larger number of pixels 
improves the image reconstruction and characterization. 
The wavelet method provides a 
better determination of $\alpha$ and assymetry as shown in Figure 8. 
Moreover, this method proves to be very promising in extracting 
the signal of low-energy events. A noticeable difference 
between the wavelet and the traditional methods can be 
observed in the results obtained from Markarian 421 and 501 data
as presented in Table 3. Moreover, the increased number of pixels 
selected by the wavelet cleaning method may
prove to be of advantage for selection methods not based on simple
moments of the light distribution, for example the 
method developed for the fine pixel camera utilized by
the CAT experiment \cite{LeBohec98}. 
Such methods do not assume simply
ellipsoidal image structure and thus can take better advantage of finer
details, at several scales, of the wavelet cleaned images.

We would like to stress the fact that
for a single telescope the limiting factor for
discriminating low energy events is still muons, with or without wavelet
cleaning.
This problem will only be solved by the construction of 
arrays of Cherenkov imaging telescopes as HEGRA \cite{hegra99} and the 
one proposed by the VERITAS project \cite{Lessard99}.
For an array of telescopes muons will not trigger the instrument.
Therefore in this case, the limiting factor will 
be the number of pixels passing the clean-up
routine and hence a wavelet image cleaning method 
as the one presented in this paper will 
be of great advantage.

\ack{The authors thank A. Haungs, S. LeBohec and T. Palfrey for 
useful comments.
This research is supported by grants from the U.S. Department of
Energy. We thank the Whipple Gamma-ray Collaboration for the use of
the data presented in this paper. L.C. thanks the Department of Physics
at Purdue University for its hospitality during her visits.}


\begin{thebibliography}{999}
\bibitem{weekes89} T.C. Weekes, et al., ApJ 342 (1989) 379.
\bibitem{bradbury99} S.M. Bradbury, et al., Proceedings of the 2xth 
	International Cosmic Ray Conference, vol 5, Salt Lake City, USA, 
	1999, p. 263.
\bibitem{lessard01} R.W. Lessard, et al., Astropart Phys 15 (2001) 1.
\bibitem{barrau98} A. Barrau et al., Nucl. Instr. and Meth. A 416 (1998) 
	278-292.
\bibitem{catanese95} M. Catanese, et al., Proceedings Towards a Major 
	Atmospheric Cherenkov Detector IV, Padova, Italy, 1995, p. 335.
\bibitem{moriarty97} P. Moriarty et al., Astropart Phys 7 (1997) 315-327.
\bibitem{hillas85} A.M. Hillas, Proceedings of 19th International Cosmic
	Ray Conference, vol. 3, La Jolla, USA, 1985, p. 445.
\bibitem{fegan94} D.J. Fegan, et al., Proceedings Towards a Major 
	Atmospheric Cherenkov Detector III, Tokyo, Japan, 1994, p. 149.
\bibitem{kertzman94} M. Kertzman, G.H. Sembroski, Nucl. Instr. and 
	Meth. A 343 (1994) 629-643.
\bibitem{mohanty98} G. Mohanty, et al., Astropart Phys 9 (1998) 15.
\bibitem{damiani97} F. Damiani, A. Maggio, G. Micela and S. Sciortino,
	ApJ 483 (1997) 350-369.
\bibitem{krywult99} J. Krywult, H.T. MacGillivray, and P. Flin, A\&A
	351 (1999) 883-892.
\bibitem{lazzati99} D. Lazzati, S. Campana, P. Rosati, M.R. Panzera and G. Tagliaferri, ApJ 524 (1999) 414-422.
\bibitem{cayon2000} L. Cay\'on, J.L. Sanz, R.B. Barreiro, E. Mart\'\i nez-Gonz\'alez, P. Vielva, L. Toffolatti, J. Silk, J.M. Diego and F. Arg\"ueso, MNRAS 315 (2000) 757-761.
\bibitem{sanz99a} J.L. Sanz, F. Arg\"ueso, L. Cay\'on, E. Mart\'\i nez-Gonz\'alez, R.B. Barreiro and L. Toffolatti, MNRAS 309 (1999a) 672-680.
\bibitem{sanz99b} J.L. Sanz, R.B. Barreiro, L. Cay\'on, E. Mart\'\i nez-Gonz\'alez, G.A. Ruiz, F.J. D\'\i az, F. Arg\"ueso, J. Silk and L. Toffolatti, A\&AS 140 (1999b) 99-105.
\bibitem{tenorio99} L. Tenorio, A.H. Jaffe, S. Hanany, and C.H. Lineweaver, 
	MNRAS 310 (1999) 823-834.
\bibitem{hobson99} M.P. Hobson, A.W. Jones, and A.N. Lasenby, MNRAS 
	309 (1999) 125-140.
\bibitem{barreiro2000} R.B. Barreiro, M.P. Hobson, A.N. Lasenby, A.J. Banday,
K.M. Gorski and G. Hinshaw, MNRAS 318 (2000) 475-481.
\bibitem{haungs99} A. Haungs, et al., Astropart Phys 12 (1999) 145-156.
\bibitem{cawley90} M.F. Cawley, et al., Exper. Astr. 1 (1990) 173.
\bibitem{reynolds93} P.T. Reynolds, et al., ApJ 404 (1993) 206.
\bibitem{punch92} M. Punch et al., Nature 358 (1992) 477.
\bibitem{quinn96} J. Quinn, et al., ApJ 456 (1996) L83.
\bibitem{LeBohec98} S. LeBohec, et al., NIM A416 (1998) 425.
\bibitem{hegra99} HEGRA Collaboration; A. Konopelko et al., Astropart Phys 10 (1999) 275.
\bibitem{Lessard99} R. Lessard, Astropart Phys 11 (1999) 243.
\end{thebibliography}
\end{document}